\documentclass[fdp,a4paper,fleqn%
]{w-art}
\usepackage{times,cite,w-thm,slashed,amssymb}
\theoremstyle{plain}

\theoremstyle{definition}

\usepackage[]{graphicx}
\numberwithin{equation}{section}
\def\Re{\mathop{\rm Re}\nolimits}
\def\Im{\mathop{\rm Im}\nolimits}
\def\rmi{{\rm i}}
\def\rmd{{\rm d}}
\def\rme{{\rm e}}
\def\cn{{\cal N}}
\newcommand{\ft}[2]{{\textstyle\frac{#1}{#2}}}
\newcommand{\Sp}{\mathop{\rm Sp}}
\newsavebox{\uuunit}
\sbox{\uuunit}
    {\setlength{\unitlength}{0.825em}
     \begin{picture}(0.6,0.7)
        \thinlines
        \put(0,0){\line(1,0){0.5}}
        \put(0.15,0){\line(0,1){0.7}}
        \put(0.35,0){\line(0,1){0.8}}
       \multiput(0.3,0.8)(-0.04,-0.02){12}{\rule{0.5pt}{0.5pt}}
     \end {picture}}
\newcommand {\unity}{\mathord{\!\usebox{\uuunit}}}

\begin{document}
\DOIsuffix{theDOIsuffix}
\pagespan{1}{}



\title[Supergravity]{Ingredients of supergravity}


\author[D. Freedman]{Daniel Z. Freedman\inst{1,}%
  \address[\inst{1}]{Department of Mathematics and Center for Theoretical Physics, M.I.T., Cambridge, Massachusetts}}
\author[A. Van Proeyen]{Antoine Van Proeyen\inst{2,}%
  \footnote{Corresponding author\quad E-mail:~\textsf{Antoine.VanProeyen@fys.kuleuven.be},
            Phone: +32\,16\,327240,
            Fax: +32\,16\,327986}}
\address[\inst{2}]{Instituut voor Theoretische Fysica, Katholieke Universiteit Leuven,\\
       Celestijnenlaan 200D B-3001 Leuven, Belgium.}

\begin{abstract}
  These notes give a summary of lectures given in Corfu in 2010 on basic
  ingredients in the study of supergravity. It also summarizes initial chapters
  of a forthcoming book `Supergravity' by the same authors.
\end{abstract}
\maketitle                   






The lectures on `Supergravity' in Corfu were inspired by a forthcoming
book \cite{bookSG}. They gave a summary of the first chapters of that
book.

The main ingredients of supersymmetry can be seen from the first
supersymmetric field theory as written in
\cite{Golfand:1971iw,Wess:1974kz}. A scalar field $A(x)$ transforms into
a fermion $\psi (x)$ with a spinor parameter $\epsilon $:
\begin{equation}
  \delta(\epsilon ) A(x)= \bar \epsilon \psi (x)\,.
 \label{delApsi}
\end{equation}
Since scalars have engineering dimension 1, and fermions have dimension
3/2, the parameter should have dimension $-1/2$. Therefore in the
transformation of the fermion into the boson, one should have a
derivative (if no negative dimension objects are introduced). The
transformation should thus be of the form
\begin{equation}
  \delta(\epsilon ) \psi (x)= \gamma ^\mu \epsilon \partial _\mu A(x)\,.
 \label{delpsiA}
\end{equation}
Details will be discussed later, but the general form leads to the idea
that the commutator of two supersymmetry transformations is a translation
\begin{equation}
  \left[ \delta (\epsilon _1),\delta (\epsilon _2)\right] = \bar \epsilon _2\gamma ^\mu \epsilon_1 \partial _\mu\,,
 \label{commdeldel}
\end{equation}
or with $Q$ the operator of supersymmetry, and $P_\mu $ the one of
translations, this gives a relation of the form $\{Q,Q\}=\gamma ^\mu
P_\mu $.

The general philosophy of the successes of field theory in the 70's was
the idea that symmetries should be promoted to local symmetries. When
this is done with the supersymmetry algebra,  the gauge theory of
translations appears and therefore the theory contains gravity, i.e.
\emph{supergravity} \cite{Freedman:1976xh}. Supergravity is a basic tool
in the study of string theory. The AdS/CFT ideas,  allow to study
non-perturbative field theories based on dualities  with supergravity
solutions. Phenomenological models in high-energy physics are developed
as supergravity theories based on compactifications on Calabi-Yau
manifolds. Also many cosmological models use a supergravity limit of
string theory. Supergravity allows also to study objects like black
holes, cosmic strings, domain walls, ..., as solutions of field equations
of local supersymmetric actions.

\section{Scalar field theory and its symmetries}

Transformations of the Poincar{\'e} group act on spacetime points as
\begin{equation}
  x^\mu = \Lambda ^\mu {}_\nu x'^\nu +a^\mu\,.
 \label{xtoxprime}
\end{equation}
We are mainly concerned with infinitesimal transformations and thus
expand the parameters of the Lorentz transformations as
\begin{equation}
  \Lambda^\mu {}_\nu =\delta ^\mu {}_\nu +\lambda^\mu {}_\nu+
 {\cal O}(\lambda ^2) \,.
 \label{infinitLambda}
\end{equation}
The generators related to the parameters $\lambda^{\mu\nu}$ satisfy the
algebra (we use the metric with signature $(-+\ldots +)$)
\begin{equation}
  [m_{[\mu\nu]},m _{[\rho\sigma]}] =
\eta_{\nu\rho}m_{[\mu\sigma]} -\eta _{\mu\rho} m_{[\nu\sigma]}
-\eta_{\nu\sigma}{}m_{[\mu\rho]} +\eta_{\mu\sigma} m_{[\nu\rho]}\,.
 \label{algebraLorentz}
\end{equation}
This algebra is satisfied by the differential operators
\begin{equation}
  L_{[\rho \sigma ]} \equiv x_\rho  \partial_\sigma- x_\sigma\partial_\rho \,.
 \label{Ldef}
\end{equation}
These are used in the definitions of the transformations of scalars using
rule $\phi (x)=\phi '(x')$:
\begin{equation}
   \phi(x) \to \phi'(x) = U(\Lambda)\phi(x) = \phi(\Lambda x+a)\,, \qquad
U(\Lambda) \equiv  \rme^{-\ft12\lambda^{\rho\sigma}L_{[\rho \sigma
]}}\,.
 \label{transformphi}
\end{equation}
For fields that are not scalars, Lorentz transformations require matrices
$m_{[\mu\nu]}$ that act on the different components and the full
transformation is
\begin{equation}
  \psi(x) \to \psi'(x) = U(\Lambda)\psi
(x) = \rme^{-\ft12\lambda^{\rho\sigma}m_{[\rho\sigma]}}\psi(\Lambda
x+a)\,.
 \label{fullPoincfields}
\end{equation}

In general, symmetries with constant parameters $\epsilon^A$ can be
written using operators $\Delta_A$, such that
\begin{equation}
  \delta \phi^i(x) \equiv \epsilon^A \Delta_A\phi^i(x)\,.
 \label{delphiDelta}
\end{equation}
The Lagrangian is not necessarily  invariant, but can transform into a
total derivative
\begin{equation}
  \delta {\cal L} \equiv \epsilon^A\left[ \frac{\delta{\cal L}}{\delta \partial_\mu\phi^i}\partial_\mu \Delta_A\phi^i +
\frac{\delta{\cal L}}{\delta\phi^i}\Delta_A\phi^i \right] =\epsilon^A \partial_\mu K^\mu_A\,.
 \label{delLdefK}
\end{equation}
This leads to conserved currents (using the Euler-Lagrange equations as
indicated by $\approx 0$)
\begin{equation}
  J^\mu{}_A =- \frac{\delta{\cal L}}{\delta \partial_\mu\phi^i} \Delta_A\phi^i + K^\mu_A\,,\qquad \partial _\mu J^\mu{}_A\approx 0\,.
 \label{currentsA}
\end{equation}
\section{The Dirac field}
The Dirac field equation is
\begin{equation}
  \slashed{\partial }
 \Psi(x) \equiv \gamma^\mu \partial_\mu \Psi(x) = m \Psi(x)\,,
 \label{Diraceqn}
\end{equation}
where the defining equation for gamma matrices is
\begin{equation}
  \{\gamma^\mu,\gamma^\nu\} \equiv \gamma^\mu
\gamma^\nu\,+\,\gamma^\nu \gamma^\mu = 2\, \eta^{\mu\nu}\,\unity\,.
 \label{gammamain}
\end{equation}
For Lorentz transformations the matrix $m_{[\mu\nu]}$ in
(\ref{fullPoincfields}), which satisfies the Lorentz algebra, is
\begin{equation}
  \Sigma^{\mu\nu} \equiv \ft{1}{4}\left[\gamma^\mu,\gamma^\nu\right] \,.
 \label{SigmaLor}
\end{equation}
\section{Clifford algebras and spinors}
We study in this chapter the gamma matrices that satisfy
(\ref{gammamain}) in various spacetime dimensions. These  determine the
properties of the spinors in the theory and of the supersymmetry algebra.
We first want to know the size of the smallest spinors in each dimension,
whether they can be chosen to be real. Furthermore we want to know which
spinor bilinears are symmetric or antisymmetric in the two spinors. The
latter is important since they will occur in any superalgebra, similar to
(\ref{commdeldel}). For the latter to be consistent we need
\begin{equation}
 \bar \epsilon _1\gamma ^\mu \epsilon _2= -\bar \epsilon _2\gamma ^\mu \epsilon _1\,.
 \label{symmetryximu}
\end{equation}

We always use gamma matrices that satisfy
\begin{equation}
  \gamma^{\mu\dagger} = \gamma^0 \gamma^\mu \gamma^0\,.
 \label{gammamudagger}
\end{equation}
Thus for spacelike $\mu $ they are Hermitian. For even dimensions $D=2m$,
we define
\begin{equation}
  \gamma_* \equiv  (-\rmi )^{m+1} \gamma_0\gamma_1\ldots\gamma_{D-1}\,,
 \label{defgammastar}
\end{equation}
which satisfies $\gamma_*^2=\unity $. E.g. for $D=4$ we have $\gamma
_*=\rmi\gamma _0\gamma _1\gamma _2\gamma _3$. These are used to split
spinors in left-handed and right-handed ones using projection operators
\begin{equation}
  P_L= \ft12 (\unity + \gamma_*)\,,\qquad P_R= \ft12 (\unity - \gamma_*)\,.
 \label{PLPR}
\end{equation}

The full Clifford algebra also contains gamma matrices that are
antisymmetric in multiple indices. They are defined by
\begin{equation}
  \gamma ^{\mu _1\ldots \mu _r}= \gamma ^{[\mu _1}\ldots \gamma ^{\mu
  _r]}\,, \qquad \mbox{e.g.}\qquad \gamma ^{\mu \nu }=\ft12\gamma^\mu \gamma^\nu
  -\ft12\gamma^\nu \gamma^\mu \,,
 \label{gammar}
\end{equation}

Since symmetries of spinor bilinears as in (\ref{symmetryximu}) are
important for supersymmetry, we use the Majorana conjugate to define
$\bar \lambda$
\begin{equation}
  \bar \lambda  \equiv \lambda  ^T C\,.
 \label{Majbar}
\end{equation}
$C$ is a matrix such that $C\gamma_{\mu _1\ldots \mu _r}$ are all
symmetric or antisymmetric, depending only on $D$ (modulo 8) and $r$
(modulo 4):
\begin{equation}
  \bar \lambda \gamma_{\mu _1\ldots \mu _r}\chi =t_r\bar \chi\gamma_{\mu _1\ldots \mu _r}
  \lambda\,, \qquad t_r=\pm 1\,.
 \label{transposetr}
\end{equation}
For general spinors (\ref{Majbar})  differs from the Dirac adjoint $\bar
\Psi\equiv \Psi ^\dagger \rmi\gamma^0$, but we will see below that it
agrees for certain types of spinors that are called Majorana spinors. The
values of $t_r$ are constrained by consistency conditions in any
dimension. E.g. for $D=2,3,4$ mod~8, which we will mainly use in these
lectures one can use $t_0=t_3=1$, $t_1=t_2=-1$. These sign factors also
determine the adjoint of composite expressions of spinors and gamma
matrices, i.e.
\begin{equation}
  \chi=\Gamma ^{(r_1)}\Gamma ^{(r_2)}\cdots \Gamma ^{(r_p)}\lambda \ \Longrightarrow
  \ \bar \chi =t_0^pt_{r_1}t_{r_2}\cdots t_{r_p}\,
  \bar\lambda \Gamma ^{(r_p)}\cdots\Gamma ^{(r_2)}\Gamma ^{(r_1)}\,.
 \label{chibar}
\end{equation}
When we need spinor indices, we contract them always in NW-SE convention,
using ${\cal C}^{\alpha \beta }$, which are the components of $C^T$ and
${\cal C}_{\alpha \beta }$, which are the components of $C^{-1}$, such
that $\lambda _\alpha $ are the components of a spinor $\lambda $ and
$\lambda ^\alpha $ are those of $\bar \lambda $:
\begin{equation}
 \lambda ^\alpha ={\cal C}^{\alpha \beta }\lambda _\beta\,,\qquad   \lambda _\alpha = \lambda ^\beta {\cal C}_{\beta \alpha }\,.
 \label{componentslambdaupdown}
\end{equation}
The components of ordinary gamma matrices are written as $(\gamma _\mu
)_\alpha {}^\beta $, whose indices can be raised or lowered with the same
rules to get to symmetric or antisymmetric matrices:
\begin{equation}
  (\gamma _\mu )_{\alpha \beta }=(\gamma _\mu )_\alpha {}^\gamma {\cal C}_{\gamma \beta
  }=-t_1(\gamma _\mu
)_{ \beta\alpha }\,.
 \label{symmgammadown}
\end{equation}

Complex conjugation can be replaced by an operation called charge
conjugation. The latter acts as complex conjugation on scalars, and has a
simple action on fermion bilinears.\footnote{In terms of complex
conjugate, the charge conjugate of a spinor is $\lambda ^C\equiv \rmi
t_0\gamma ^0C^{-1}\lambda ^*$.} For example, it preserves the order of
spinor factors. For all practical purposes one can consider the charge
conjugate $\lambda ^C$ as the complex conjugate of the spinor $\lambda $.
For a spinor bilinear, using a matrix in spinor space $M$, we have
\begin{equation}
  \left( \bar \chi M\lambda \right) ^*\equiv \left( \bar \chi M\lambda \right)
  ^C= (-t_0t_1)\overline{ \chi ^C}M^C\lambda^C\,.
 \label{defcc}
\end{equation}
Thus one only has to know the charge conjugate of matrices in spinor
space, e.g.
\begin{equation}
  (\gamma _\mu )^C= (-t_0t_1)\gamma _\mu \,,\qquad  (\gamma _*)^C = (-)^{D/2+1}\gamma _*\,.
 \label{gammaC}
\end{equation}

A priori a spinor has $2^{\mathop{\rm Int}\nolimits[D/2]}$ (complex)
components. We saw already that for even dimensions they can be reduced
by a factor 2 using the projections (\ref{PLPR}). These define \emph{Weyl
spinors}. In some dimensions (and spacetime signature, but we will always
assume Minkowski signature here) there are reality conditions $\psi =\psi
^C$, consistent with Lorentz algebra. This defines \emph{Majorana
spinors}. The consistency condition for this definition can be expressed
in terms of the sign factors in (\ref{transposetr}) as $t_1 = -1$. With
(\ref{PLPR}) and (\ref{gammaC}) it is easy to see that such a reality
condition can only be consistent with a chiral projection if $D=2$ mod 4
(and due to $t_1=-1$ in fact $D=2$ mod 8). When we can define real chiral
spinors, we call them \emph{Majorana-Weyl}. This leads in $D=10$ to
spinors with only 16 real components. In $D=4$ both Majorana spinors and
Weyl spinors have 4 real components. It is equivalent to write fermions
in terms of Majorana spinors $\psi $ or in terms of the Weyl spinors
$P_L\psi $:
\begin{equation}
  (P_{L}\psi)^C=  P_R\psi\,,\qquad (P_{R}\psi)^C=  P_L\psi\,.
 \label{PLpsiC}
\end{equation}
In dimensions with $t_1=1$ one can define reality conditions for doublets
of spinors
\begin{equation}
  \chi^i=
 \varepsilon ^{ij}(\chi ^j)^C\,.
 \label{symplM}
\end{equation}
This defines \emph{symplectic Majorana-Weyl} spinors. But this does in
fact not diminish the minimal number of real components. E.g. for $D=5$
the minimal spinor has either as Dirac spinor or as symplectic-Weyl
spinor 8 real components.

\section{The Maxwell and Yang-Mills Gauge Fields}

Maxwell fields with gauge transformations $\delta A_\mu=\partial _\mu
\theta (x)$ couple typically to complex fields like spinors that
transform like $\delta \psi = \rmi q\theta \psi $. The standard actions
are then of the form
\begin{eqnarray}
  S[A_\mu,\bar{\Psi},\Psi] &=&\int \rmd^Dx\left[
-\ft{1}{4}F^{\mu\nu}F_{\mu\nu} -\bar{\Psi}(\gamma^\mu D_\mu
-m)\Psi\right]\,,\nonumber\\
&& D_\mu \Psi \equiv (\partial_\mu - \rmi q A_\mu)\Psi\,,\qquad F_{\mu\nu} \equiv \partial_\mu A_\nu -\partial_\nu A_\mu\,.
 \label{typactAbelian}
\end{eqnarray}

In many supersymmetric theories several Abelian gauge vectors
$A_\mu{}^A$, $A=1,\ldots ,m$, appear and the generalized electromagnetic
duality transformations play an important role. These apply in general
actions for $D=4$ of the form
\begin{equation}
 {\cal L} = -\ft14(\Re f_{AB})
F_{\mu \nu }^A F^{\mu \nu \,B} +\ft18\varepsilon ^{\mu \nu \rho \sigma }(\Im f_{AB}) F_{\mu \nu }^A
F_{\rho \sigma }^B\,,\quad\varepsilon_{012(D-1)}=1\,,\quad \varepsilon^{012(D-1)}=-1\,,
 \end{equation}
where $f_{AB}$ is a complex matrix that may depend on scalar fields in
the theory. Using notations with (anti)self-dual tensors
\begin{equation}
{F}^{\pm A }_{\mu\nu}\equiv \ft12\left( {F}^A_{\mu\nu}
\pm \tilde {F}_{\mu\nu}^A\right)\,,\qquad\tilde F^{\mu \nu}=-\ft12\rmi\varepsilon ^{\mu \nu \rho \sigma }{F}_{\rho \sigma
}^A\,,
 \label{Fpm}
\end{equation}
the Bianchi identities and field equations can be written in a similar
form:
\begin{equation}
  \partial^\mu \Im {F}^{A\,- }_{\mu\nu} =0\,,\qquad \partial_\mu \Im G_{A }^{\mu\nu\,-}=0\,,
 \label{BianchiFE}
\end{equation}
where
\begin{equation}
\label{gtensor} G^{\mu\nu}_A \equiv
\varepsilon^{\mu\nu\rho\sigma}\frac{\delta  S}{\delta
F^{\rho\sigma\,A}}\,,\qquad   G_A^{\mu \nu \,-}=\rmi f_{AB}F^{\mu \nu  \,-\,B}\,.
\end{equation}
A priori, the equations (\ref{BianchiFE}) seem to be invariant under
general real linear transformations
\begin{equation}
\begin{pmatrix}
 F'^-\cr  G'^-
\end{pmatrix}
={\mathcal S}
\begin{pmatrix}
 F^-\cr G^-
\end{pmatrix}
\equiv
\begin{pmatrix}
 A&B\cr C&D
 \end{pmatrix}
\begin{pmatrix}
 F^-\cr G^-
 \end{pmatrix}
\,.
\end{equation}
The  compatibility with (\ref{gtensor}) requires that ${\cal S}$ is a
symplectic matrix, and that simultaneously the matrix $f_{AB}$ transforms
as
\begin{equation}
  \rmi f' = (C + \rmi Df)(A+\rmi Bf)^{-1}\,.
 \label{transfof}
\end{equation}
We thus conclude that the duality transformations in 4 dimensions are
transformations in the symplectic group $\Sp (2m,\mathbb R)$.
\cite{Gaillard:1981rj}

\section{The free Rarita-Schwinger field}
Massless spin 3/2 fields are described by spinor-vectors $\Psi_{\mu}(x)$
that transform with a local spinorial parameter as $\delta \Psi_{\mu}(x)=
\partial_\mu \epsilon(x)$. The action that gives the right propagating
degrees of freedom is
\begin{equation}
  S = -\int \rmd^Dx\, \bar{\Psi}_\mu\gamma^{\mu\nu\rho}\partial_\nu\Psi_\rho\,.
 \label{masslessRS}
\end{equation}
Its field equation, $\gamma^{\mu\nu\rho}\partial_\nu\Psi _\rho=0$ is
equivalent to
\begin{equation}
  \gamma^\mu(\partial_\mu \Psi_\nu - \partial_\nu\Psi_\mu) =0\,,
 \label{equivRSeqn}
\end{equation}
and after choosing a gauge fixing leads to solutions in terms of
$(D-3)2^{[{D\over 2}]}$ initial conditions.

It is useful to remind that there are two countings of the number of
degrees of freedom (dof) of fields:
\begin{description}
  \item[on-shell counting] is the number of helicity states, and can be
  obtained by counting the number of independent initial conditions
  for the field equations divided by~2 (since these are coordinates
  and momenta).
  \item[off-shell counting] is the number of field components $-$ the
  number of gauge transformations.
\end{description}
While real scalars have 1 dof in both countings, a real spinor has
$2^{[D/2]}$ off-shell, and half of these on-shell dof. A massless vector
has $D-1$ off-shell and $D-2$ on-shell dof. For the massless $\Psi_\mu $
field we find here $(D-1)2^{[{D\over 2}]}$ off-shell and
$\ft12(D-3)2^{[{D\over 2}]}$ on-shell dof. The graviton field has
$\ft12D(D-1)$ off-shell and $\ft12D(D-3)$ on-shell dof.

\section{$\cn =1$ global supersymmetry in $D=4$}
The (classical) supersymmetry algebra is the completion of the Poincar{\'e}
group with spinorial generators $Q_\alpha $ that satisfy anticommutation
relations
\begin{equation}
  \left\{Q_\alpha,Q_\beta\right\} =-\ft12(\gamma^\mu)_{\alpha\beta} P_\mu\,.
 \label{SUSYQQ}
\end{equation}
The supersymmetry transformations with constant spinor parameters can be
written as $\delta(\epsilon ) =\bar \epsilon Q=\epsilon ^\alpha
Q_\alpha$, and $P_\mu $ is classically realized on fields as $\partial
_\mu $. A basic realization of this algebra is the chiral multiplet with
complex scalar fields $Z$ and $F$, and Majorana spinor $\chi $ (or its
chiral projection $P_L\chi $):
\begin{equation}
  \delta(\epsilon ) Z = \frac1{\sqrt{2}}\bar{\epsilon} P_L \chi\,, \qquad
\delta(\epsilon ) P_L\chi = \frac1{\sqrt{2}} P_L(\slashed{\partial} Z +
F)\epsilon\,,\qquad
 \delta(\epsilon ) F = \frac1{\sqrt{2}} \bar{\epsilon}\, \slashed{\partial}
P_L\chi\,.
 \label{chiralmult}
\end{equation}
The simplest actions are those with kinetic terms and potential terms:
\begin{eqnarray}
  S &=& S_{\rm kin} +  S_F + S_{\bar{F}}\,,\label{Schiralm}\\
  S_{\rm kin} &=& \int {\rm d}^4x[-\partial ^\mu \bar{Z}\partial _\mu Z
-\bar{\chi} \slashed{\partial } P_L\chi +\bar{F}F]\,,\qquad
S_F = \int {\rm d}^4x [FW'(Z) - \ft12
\bar{\chi}P_LW''(Z)\chi]\nonumber\,,
\end{eqnarray}
where $F(Z)$ is a holomorphic function, called the superpotential.

We present the `gauge multiplet'  for non-Abelian gauge fields. The
action, supersymmetry and gauge transformations are
\begin{eqnarray} S_{\rm gauge} &=& \int \rmd ^4x\,\left[ -\ft{1}{4}F^{\mu\nu A}F^A_{\mu\nu} -\ft12
\bar{\lambda}^A \gamma^\mu D_\mu \lambda^A +\ft12 D^AD^A\right]\,,\nonumber\\
\delta A^A_\mu &=& -\ft{1}{2}\bar{\epsilon}\gamma_\mu \lambda^A+\partial_\mu\theta^A +\theta^C A_\mu{}^B
f_{BC}{}^A\,,\nonumber\\
\delta\lambda^A &=& \left[\ft{1}{4} \gamma^{\mu \nu }F^A_{\mu \nu }
+\ft{1}{2}\rmi \gamma_*
D^A\right]\epsilon+\theta^C \lambda ^B
f_{BC}{}^A\,,\\
 \delta D^A &=& \ft{1}{2}\rmi\,\bar{\epsilon} \gamma_* \gamma^\mu D_\mu \lambda^A+\theta^C D ^B
f_{BC}{}^A\,, \qquad
 D_\mu \lambda ^A\equiv \partial_\mu \lambda ^A +\lambda ^C A_\mu{}^B
 f_{BC}{}^A\,.\nonumber
\end{eqnarray}
$f_{BC}{}^A$ are the structure constants of the gauge group, which
commutes with supersymmetry, but enters in the commutator of two
supersymmetries:
\begin{equation}
  \left[\delta(\epsilon _1),\delta(\epsilon _2)\right]= -\ft12 \bar{\epsilon}_1\gamma^\nu\epsilon_2\partial _\nu
  + \delta \left(\theta ^A=\ft12 \bar{\epsilon}_1\gamma^\nu\epsilon_2A _\nu{}^A\right)\,.
 \label{commsusygauge}
\end{equation}
The right-hand side is called a `gauge-covariant translation'. The gauge
multiplet can be coupled to chiral multiplets transforming in a
representation of the gauge group. In that case, some supersymmetry
transformations of the chiral multiplet are modified:
\begin{equation}
\delta(\epsilon ) P_L\chi = \frac1{\sqrt{2}} P_L(\slashed{D} Z +
F)\epsilon\,,\qquad
 \delta(\epsilon ) F = \frac1{\sqrt{2}} \bar{\epsilon}\, \slashed{D}
P_L\chi-\bar{\epsilon}P_R\lambda^A t_A Z\,,
 \label{chiralmultgauge}
\end{equation}
where $\theta ^At_A Z$ is the gauge transformation of the scalar field,
and $D_\mu=\partial _\mu -A_\mu {}^At_A $ are the gauge-covariant
derivatives. Also in the action (\ref{Schiralm}) ordinary derivatives are
replaced by covariant derivatives, and supersymmetry requires an extra
coupling term
\begin{equation}
  S_{\rm coupling} = \int \rmd ^4x 
\left[ -\sqrt{2}(\bar{\lambda}^A \bar{Z}t_A P_L\chi
-\bar{\chi}P_R t_A Z \lambda^A) +
\rmi\,D^A \bar{Z}t_A Z\right]\,.
 \label{Scoupling}
\end{equation}

\section{Differential geometry}

For gravity we have to provide spacetime with a possibly nontrivial
metric $g_{\mu \nu }(x)$. In any point of spacetime, this can be brought
to a standard form using a `frame field' $e_\mu ^a(x)$:
\begin{equation}
  g_{\mu \nu}(x)=e_\mu ^a(x)\eta _{ab}e_\nu ^b(x)\,.
 \label{gmunuemua}
\end{equation}
In gravity theories with fermions, we have to make use of these frame
fields. The Levi-Civita alternating symbol written in one or the other
indices are related by
\begin{eqnarray}
\varepsilon_{\mu_1 \mu_2 \cdots \mu_D} &\equiv&e\,^{-1} \varepsilon_{a_1 a_2
\cdots a_D}e^{a_1}_{\mu_1} e^{a_2}_{\mu_2}\cdots e^{a_D}_{\mu_D}\,,\nonumber\\
\varepsilon^{\mu_1 \mu_2\cdots \mu_D} &\equiv&e\, \varepsilon^{a_1 a_2\cdots
a_D}e_{a_1}^{\mu_1} e_{a_2}^{\mu_2}\cdots e_{a_D}^{\mu_D}\,,
 \end{eqnarray}
where $e=\det e_\mu ^a$, while for all other vectors or tensors the
components are related by equations of the type $V_\mu =e_\mu ^aV_a$.

Differential forms offer often a convenient way to write field theories.
We define the connection between components, $p$-forms and their exterior
derivatives by
\begin{equation}
  \omega^{(p)} = \frac{1}{p!}\omega_{\mu_1\mu_2\cdots\mu_p}\rmd
x^{\mu_1}\wedge \rmd x^{\mu_2}\wedge \ldots \rmd x^{\mu_p}\,,
\quad  \rmd\omega^{(p)} = \frac{1}{p!}
\partial_\mu\omega_{\mu_1\mu_2\cdots\mu_p}\rmd x^\mu \wedge \rmd
x^{\mu_1}\wedge \rmd x^{\mu_2}\wedge \ldots \rmd x^{\mu_p}\,.
 \label{diffforms}
\end{equation}
The frame fields allow to write a local basis of 1-forms $e^a \equiv
e^a_\mu(x) \rmd x^\mu$. These are used to define the Hodge duals of
$p$-forms as $(D-p)$-forms
\begin{equation}
   ^*e^{a_1}\wedge\ldots e^{a_p} = \frac{1}{q!}e^{b_1}\wedge\ldots e^{b_q}\varepsilon_{b_1 \cdots b_q}{}^{a_1 \cdots
a_p} \,.
 \label{Hodgedualdef}
\end{equation}
These definitions lead to the expression
\begin{equation}
  \int{}^*\omega^{(p)}\wedge \omega^{(p)} = \frac{1}{p!}\int \rmd^D x\,
\sqrt{-g}\, \omega^{\mu_1 \cdots \mu_p} \omega_{\mu_1\cdots \mu_p}\,,
 \label{intomstarom}
\end{equation}
which can be used to write a generalization of the Maxwell action as
\begin{equation}
  S_p= -\ft12\int {}^*F^{(p+1)} \wedge F^{(p+1)} \,,\qquad F^{(p+1)}\equiv\rmd A^{(p)}\,.
 \label{Sp}
\end{equation}
The $p=1$ case is the Maxwell field. The field equations and Bianchi
identities for $F^{p+1}$ can be interpreted as Bianchi identities and
field equations for ${}^*F^{p+1}$, which can therefore be interpreted as
the field strength of a $(D-p-2)$-form. This implies that such actions
for $p$-forms and for $(D-p-2)$-forms are equivalent. This is the
generalization of the electric-magnetic duality of Maxwell fields in 4
dimensions, where the electric 1-form is transformed to a magnetic
1-form. Further, for 4 dimensions it implies that an action (\ref{Sp})
for an antisymmetric tensor (2-form) is equivalent to a scalar action. In
higher dimensions, this is, however, not the case, and 3-forms are
essential to construct $D=11$ supergravity.

For supergravity one must use covariant derivatives. They involve
connections that are gauge fields for the Lorentz transformations, i.e.
$\omega _\mu{} ^{ab}=-\omega_\mu{} ^{ba}$ such that under the
infinitesimal transformations of (\ref{infinitLambda}), $\delta\omega
_\mu{} ^{ab}=\partial  _\mu{} \lambda ^{ab}-
\lambda^a{}_c\omega_\mu{}^{cb} +\omega_\mu{}^{ac}\lambda_c{}^b$, as
appropriate for a gauge field for the algebra (\ref{algebraLorentz}).
Furthermore, there is an affine connection $\Gamma _{\mu \nu }^\rho $
determined by the \emph{vielbein postulate}
\begin{equation}
 \nabla _\mu e_\nu ^a=\partial_\mu e^a_\nu +\omega_\mu {}^a{}_b e^b_\nu -\Gamma^\sigma_{\mu\nu}e^a_\sigma = 0\,.
 \label{vielbeinpost}
\end{equation}
`Torsion' is the antisymmetric part of $\Gamma $. In form language
\begin{equation}
  \rmd e^a + \omega^a{}_b \wedge e^b \equiv T^a=\ft12T_{\mu \nu }{}^a\rmd x^\mu \wedge \rmd x^\nu \,,\qquad \Gamma^\rho_{\mu\nu}-\Gamma^\rho_{\nu\mu}=T_{\mu\nu}{}^\rho\,.
 \label{torsionform}
\end{equation}
We can split these connections in the metric-part and the
torsion-dependent part:
\begin{eqnarray} \omega_\mu{}^{ab} &=& \omega_\mu{}^{ab}(e) + K_\mu{}^{ab}\,,\qquad
\omega _\mu {}^{ab}(e)= 2 e^{\nu[a} \partial_{[\mu} e_{\nu]}{}^{b]} -
e^{\nu[a}e^{b]\sigma} e_{\mu c} \partial_\nu e_\sigma{}^c\,,\nonumber\\
&&\phantom{\omega_\mu{}^{ab}(e) + K_\mu{}^{ab}\,,\qquad}
 K_{\mu[\nu\rho]} = -\ft12 (T_{[\mu\nu]\rho}
-T_{[\nu\rho]\mu}+T_{[\rho\mu]\nu})\,,\nonumber\\
\Gamma^\rho_{\mu\nu}&=& \Gamma^\rho_{\mu\nu}(g) -K_{\mu\nu }{}^\rho \,,\qquad
 \Gamma^\rho_{\mu\nu}(g)= \ft12 g^{\rho\sigma}(\partial_\mu
g_{\sigma\nu}+\partial_\nu g_{\mu\sigma}-\partial_\sigma g_{\mu\nu})\,. \label{omegaGamma}
\end{eqnarray}
The curvature tensor can be defined from both connections, and an
integrability condition on (\ref{vielbeinpost}) implies that these are
related by changing indices in the conventional way:
\begin{eqnarray}
&& R_{\mu\nu ab} \equiv \partial_\mu \omega_{\nu ab} -\partial_\nu
\omega_{\mu ab} +\omega_{\mu ac}\omega_\nu {}^c{}_b -\omega _{\nu
ac}\omega_\mu {}^c{}_b\,,\nonumber\\
&& R_{\mu\nu}{}^\rho{}_\sigma \equiv \partial_\mu \Gamma^\rho_{\nu\sigma}
-\partial_\nu \Gamma^\rho_{\mu\sigma} +
\Gamma^\rho_{\mu\tau}\Gamma^\tau_{\nu\sigma} -
\Gamma^\rho_{\nu\tau}\Gamma^\tau_{\mu\sigma}\,,\qquad
R_{\mu\nu}{}^\rho{}_\sigma = R_{\mu\nu ab} e^{a\rho} e^b_\sigma \,.
 \label{curvature}
\end{eqnarray}
\section{The first and second order formulations of general relativity}
General relativity is well-known for bosonic fields, but there are
several subtleties when fermions are involved. Let us consider
\begin{equation}
 S = S_2+S_{1/2}=\int \rmd^Dx \,e\,  \left[\frac{1}{2\kappa^2}
e_a^{\mu}e_b^{\nu} R_{\mu\nu}{}^{ ab} - \frac12\bar{\Psi}\gamma^\mu
\nabla_\mu\Psi +{1\over 2}\bar{\Psi} {\raise.3ex\hbox{$\stackrel{\leftarrow}{\nabla}$}{}}_\mu \gamma^\mu\Psi\right]\,,
\label{S212}
\end{equation}
where $\kappa $ is the gravitational coupling constant, $\gamma^\mu =
e_a^\mu \gamma^a$ and
\begin{equation}
\nabla_\mu \Psi  = (\partial_\mu + \ft{1}{4}\omega_{\mu}{}^{ab}\gamma_{ab})\Psi\,,\qquad
\bar{\Psi} {\raise.3ex\hbox{$\stackrel{\leftarrow}{\nabla}$}{}}_\mu= \bar{\Psi}( \raise.3ex\hbox{$\stackrel{\leftarrow}{\partial }$}{}_\mu
-\ft{1}{4}\omega_{\mu}{}^{ab} \gamma_{ab})\,,
 \label{LorcovderPsi}
\end{equation}
are the Lorentz covariant derivatives. We consider here first that
$\omega_{\mu}{}^{ab}$ is only $\omega_{\mu}{}^{ab}(e)$, i.e. without
torsion, and $R_{\mu\nu}{}^{ ab}$ is the expression (\ref{curvature})
using $\omega_{\mu}{}^{ab}(e)$. The derivatives in (\ref{curvature}) and
(\ref{omegaGamma}) imply that the action is a second order action in the
independent field $e_\mu ^a$. Therefore, this is called `\emph{second
order formalism}'.

Another approach considers $e_a^\mu$ and $\omega_{\mu}{}^{ab}$ as
independent fields. The action is then only first order in independent
fields, and therefore this is called `\emph{first order formalism}'. We
consider first the field equation for $\omega_{\mu}{}^{ab}$. Without the
fermionic part this would give
$\omega_{\mu}{}^{ab}=\omega_{\mu}{}^{ab}(e)$. Therefore, comparing with
(\ref{omegaGamma}) shows that the fermionic terms determine torsion. In
fact, we get the solution as in (\ref{omegaGamma}) with
\begin{equation}
 \omega_\mu{}^{ab} = \omega_\mu{}^{ab}(e) + K_\mu{}^{ab}\,,\qquad
K^\nu{}_{ab} = -\ft14\kappa^2 \bar{\Psi}\,\gamma_{ab}{}^\nu\,\Psi \,, \qquad
 T_{ab}{}^\nu= -2K^\nu{}_{ab}\,.
 \label{Torsionfermion}
\end{equation}
Thus, using this result, we have the same action as in (\ref{S212}), but
with another definition of $\omega_{\mu}{}^{ab}$. To make the difference
explicit, one may rewrite that action by expanding (\ref{Torsionfermion})
and this gives
\begin{equation}
   S= 
S (T=0)
+\frac1{32} \kappa^2 \int \rmd^Dx \,e\,\,(\bar{\Psi}\gamma_{\mu\nu\rho}\Psi)(
\bar{\Psi}\gamma^{\mu\nu\rho}\Psi)\,,
 \label{S4ferm}
\end{equation}
where the first term is (\ref{S212}) with the torsionless connection. Due
to the smallness of the gravitational coupling constant, the 4-fermion
contact term cannot be measured in practice, but in principle there is a
physical difference between this first and second order action. To obtain
local supersymmetric actions, it turns out that in supergravity similar
terms appear which can be understood from the first-order formalism
\cite{Deser:1976eh}.

\section{Outlook}

Supergravity is based on the ingredients mentioned in these lectures. In
fact, the reader is  now well equipped to study the first supergravity
theories.

\begin{acknowledgement}
We thank the organizers of the school for the nice atmosphere and the
opportunity to teach and receive feedback from students during these
lectures.

This work is supported in part by the FWO - Vlaanderen, Project No.
G.0651.11, and in part by the Federal Office for Scientific, Technical
and Cultural Affairs through the ``Interuniversity Attraction Poles
Programme -- Belgian Science Policy'' P6/11-P.
\end{acknowledgement}


\providecommand{\WileyBibTextsc}{}
\let\textsc\WileyBibTextsc
\providecommand{\othercit}{} \providecommand{\jr}[1]{#1}
\providecommand{\etal}{~et~al.}

\end{document}